*ISSI chapter: Water and the interior structure of terrestrial planets and icy bodies*

J. Monteux[1], G. J. Golabek[2], D. C. Rubie[2], G. Tobie[3] & E. D. Young[4]

[1] Laboratoire Magmas et Volcans, Université Blaise Pascal Clermont-Ferrand, 6 Avenue Blaise Pascal, 63178 Aubiere Cedex, France.

[2] Bayerisches Geoinstitut, Universität Bayreuth, Universitätsstrasse 30, 95440 Bayreuth, Germany.

[3] Laboratoire de Planétologie et Géodynamique, UMR-CNRS 6112, Université de Nantes, 44322 Nantes Cedex, France.

[4] Department of Earth, Planetary, and Space Sciences, University of California Los Angeles, Los Angeles, CA 90095, USA.

**Abstract**

Water content and the internal evolution of terrestrial planets and icy bodies are closely linked. The distribution of water in planetary systems is controlled by the temperature structure in the protoplanetary disk and dynamics and migration of planetesimals and planetary embryos. This results in the formation of planetesimals and planetary embryos with a great variety of compositions, water contents and degrees of oxidation. The internal evolution and especially the formation time of planetesimals relative to the timescale of radiogenic heating by short-lived $^{26}$Al decay may govern the amount of hydrous silicates and leftover rock-ice mixtures available in the late stages of their evolution. In turn, water content may affect the early internal evolution of the planetesimals and in particular metal-silicate separation processes. Moreover, water content may contribute to an increase of oxygen fugacity and thus affect the concentrations of siderophile elements within the silicate reservoirs of Solar System objects. Finally, the water content strongly influences the differentiation rate of the icy moons, controls their internal evolution and governs the alteration processes occurring in their deep interiors.

**Introduction**

The Solar System displays a strong dichotomy between the inner region that is characterized by relatively dry planetary objects having a very small water fraction (<0.1 %), and the outer solar system where water ice constitutes a large fraction of solid material (> 20%), which is inherited from accretion processes. When compared with other planetary systems, the Solar System seems however rather unusual. Exoplanet surveys have revealed that planets intermediate in mass between Earth and Neptune are surprisingly common, but are notably absent in the Solar System (*Mayor et al. 2014, Howard 2012, Marcy et al. 2014*). Model mass-radius relationships indicate a great diversity of interior composition and atmospheric extent for the Super-Earth/Mini-Neptune-planet class (e.g. *Howard 2012*), suggesting a wide range of volatile contents (including water) and compositions.

The distribution of water in the planetary system is controlled by the temperature structure in the protoplanetary disk and dynamics and migration of planetesimals and planetary embryos (*Raymond et al. 2004; Cowan and Abbot, 2014; O'Brien et al., 2014; Rubie et al., 2015a*). This results in the formation of planetesimals and planetary embryos with a great variety of compositions and water contents. Subsequent accretionary processes can lead to the formation of planets that range from completely dry planets to "water worlds" with more than 100 Earth-oceans of water (1 ocean = $1.5 \times 10^{21}$ kg $H_2O$) (*Raymond et al. 2004; Cowan and Abbot, 2014; O'Brien et al., 2014; Rubie et al., 2015a*). As the evolution of water content inside a growing proto-planet is a strong function of pressure-temperature conditions, the early stages of planetary formation that govern their accretion rates and their early heat budget may have played a major role in the water distribution in the solar system. Indeed, depending on the accretion rate and on the formation time of the building blocks of the terrestrial planets relative to the timescale of radiogenic heating by the decay of short-lived $^{26}$Al, water can either be incorporated as an ice-rock mixture or in the form of hydrated silicates or be absent. Aqueous alteration appears to be common in carbonaceous chondrites (*Kerridge and Bunch, 1979; Zolensky et al. 1989; Krot et al. 1998*). The proportion of planetesimal water incorporated into rock depends critically on whether water was mobile or stagnant (e.g., *Young et al., 1999; Young 2001; Young et al. 2003; Bland et al. 2009; Fu et al. 2017; Castillo-Rogez and Young, 2017*). In turn, the water content of Solar System objects likely played a key role in the internal evolution of the solar system bodies. For example, water dissolved in silicate minerals is known to significantly reduce mantle viscosity and melting temperature (e.g. *Hirth and Kohlstedt, 1996; Katz et al., 2003; Hirschmann, 2006*). It has therefore a potential influence on the early differentiation processes and a strong feedback on the thermal state of the mantle and the vigor of convection (e.g. *Korenaga, 2010*).

Water is also the major constituent of the moons orbiting Jupiter (except for Io), Saturn, Uranus and Neptune, and all planetary objects beyond Neptune (e.g. *Hussmann et al., 2015*). In these icy objects, water is likely present either as thick ice shells or deep water oceans (e.g. *Hussmann et al., 2006*) and in some cases is being ejected from venting plumes composed mainly of water vapor and ice particles

(e.g. *Porco et al., 2006* for Enceladus, *Roth et al., 2014* for Europa). The combined effect of accretional heating, radiogenic decay by short-lived isotopes, tidal heating associated with despinning, and viscous heating associated with sinking negative rock diapirs may increase the internal temperature above the melting point of ice (e.g. *Tobie et al. 2013*). This may result in a partial to full ice-rock separation. During this differentiation as well as at the present-day, interactions between solid rock and liquid water might have promoted chemical reactions facilitated by the presence of ammonia (*Schubert et al., 2010; Sohl et al., 2010*). It is also possible that aqueous alteration might have led to serpentinization of olivine-rich rocks and to the formation of a highly hydrated rocky core that may result from a rapid differentiation process that implies large-scale melting events, as proposed for Saturn's moon Titan (e.g. *Castillo-Rogez and Lunine, 2010*).

Finally, comets and comet-like objects are another class of objects where ice has been detected as a major component (80%) (*Encrenaz, 2008*). These bodies, with radii usually smaller than 10 km, were formed in the outer Solar System. Comets are undifferentiated and, hence, represent weakly altered remnants of the early stages of solar-system formation. Because of radiogenic heating from $^{26}$Al and $^{60}$Fe, the interior of comets may reach temperatures above the melting point of water and ultravolatile species may be lost (*Prialnik et al., 2008, Mousis et al., 2017*). In the following study, we do not consider the comets and we will only focus on objects with radii larger than 50 km (see review of *Prialnik et al., 2008* for more details).

The structure of this article is as follows: We first detail the distribution of water in the protoplanetary disc and its effect on oxidation. Second, we present a discussion of the effect of internal evolution of planetesimals on water content based on numerical models. In particular we investigate the influence of planetesimal radius and formation time. We then discuss the role of water on the internal evolution of proto-planets and terrestrial bodies focusing particularly on metal/silicate separation and oxygen fugacity evolution. Finally we briefly describe the internal evolution of icy satellites from their accretion to their present-day activity and emphasize the role of water in differentiation and alteration processes.

## 1. Distribution of water in the protoplanetary disc and its effect on oxidation.

Protoplanetary disks of gas and dust are by necessity characterized by radial increases in temperature and pressure towards the nascent star at their center. The inner portions of the disk are heated by two mechanisms. So-called "passive" disks are heated at their surfaces by illumination from the central star. Viscous heating in "active" accretion disks can warm up the inner portions of the disk to temperatures well above those imposed by passive illumination alone. The degree of viscous heating depends on the mechanism for angular transport (accretion onto the star) (e.g., *Bai and Stone, 2013*). The greater the effective viscosity of the disk, the greater the heating related to accretion. In all cases, there will be specific radii corresponding to the condensation fronts for various volatile components (referred to as "snowlines"). From the perspective of volatiles and planet formation, the most important of these may be that for water. The water snowline separating the inner regions of the disk where water vapor is stable from the more distal regions of the disk where water ice is stable is thought to exert a first-order control on the chemistry of planetesimals. The snowline in an accretion disk is closest to the star at the midplane where the intervening mass of gas and dust shields the disk from light from the central star. Higher in the disk, the snowline curves to greater radial distances due to heating of the surfaces of the disk by the central star. We are most concerned with the position of the water snowline at the midplane when contemplating planetesimal formation. However it must be kept in mind that the time-dependent position of the snowline in the midplane and its effects on planetesimal formation over time are still debated (e.g., *Morbidelli et al., 2016*).

The bulk compositions of planetesimals and planetary embryos in the protoplanetary disc, including water and some volatile element abundances, have been constrained by core formation modelling (Rubie et al., 2015a, 2016). Mantle abundances of siderophile elements (Ni, Co, W etc.) can be reproduced in multistage core formation models when the initial 60-70% of Earth accretes from highly reduced material and the final 30-40% from relatively-oxidized material, in combination with the effects of Si partitioning into the core (*Rubie et al., 2011*). The late accretion of oxidized material is also consistent with the relatively late accretion of volatile elements to the Earth (e.g. *Schönbächler et al., 2010; Dauphas, 2017*). This scenario was developed further by *Rubie et al. (2015a, 2016)* by formulating a combined accretion/core formation model. This was based on "Grand Tack" N-body accretion models (*Walsh et al., 2011; O'Brien et al., 2014*) that started with up to ~200 embryos (distributed from 0.7 to 3.0 au) embedded in a protoplanetary disc consisting of up to 4400 planetesimals (distributed from 0.7 to 9.5 au). Each accretional collision between embryos and other embryos and planetesimals was considered to result in an episode of core formation, provided the impactor was metal-bearing. The bulk composition of each of the starting bodies was defined in terms of CI ratios of non-volatile elements, with oxygen and water contents varying as a function of heliocentric distance. Based on these bulk compositions, the compositions of equilibrated metal and silicate in each core-formation event were determined by a combined mass balance/element partitioning approach, thus enabling the evolving compositions of the mantles and cores of all growing planets to be modeled. The composition of the model Earth in each simulation is fit to Earth's mantle composition by a least squares refinement of the metal-silicate equilibration pressure and several parameters that describe the compositions of initial bodies (Fig. 1).

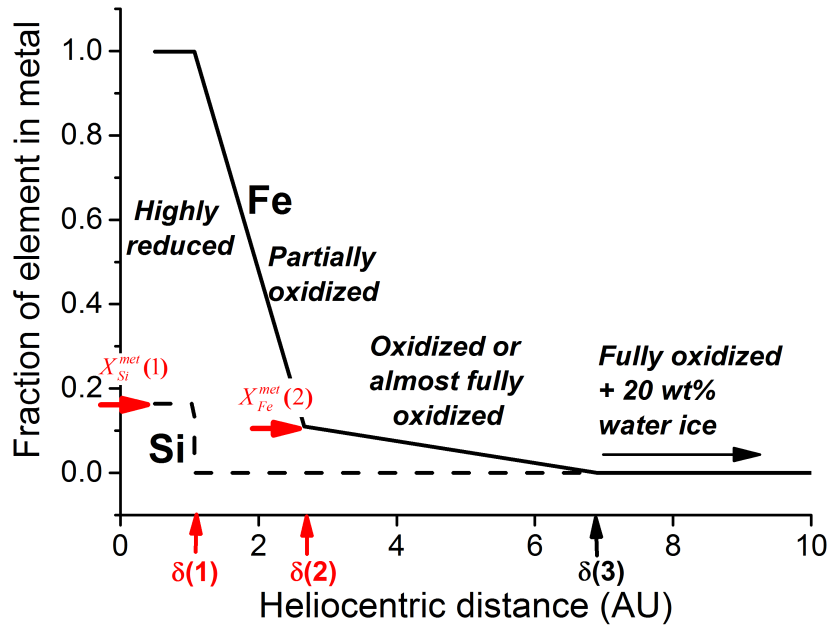

Fig. 1. The composition of initial bodies (embryos and planetesimals) as a function of heliocentric distance determined in the accretion/core formation model of Rubie et al. (2015a). At heliocentric distances less than ~1 au, 99% of all Fe is present as metal and ~18% of total Si is dissolved in the metal – which makes the bulk composition very reduced. Beyond ~1 au, compositions become increasingly oxidized with increasing heliocentric distance because the fraction of Fe present as metal decreases and the fraction of Fe present as FeO in the silicate increases. Beyond 6-7 au, compositions are fully oxidized (no metal) and contain 20 wt% water – these are bodies that originally accreted beyond the snowline. The red arrows indicate values of parameters that were refined by least squares.

Best fits to Earth's mantle composition were obtained when compositions of bodies close to the Sun are highly reduced and those further out are relatively oxidized (Fig. 1).

The gradient in Fig. 1 requires explanation in the context of the young solar protoplanetary disk. Dynamical models have shown that radial mixing within the protoplanetary disk was likely to occur especially during the late stages of planetary formation (e.g. *Chambers, 2001*; *Raymond et al., 2004*). Variation in oxygen fugacity across the protoplanetary gaseous disk is a longstanding question. At temperatures greater than about 500 to 600 K or so, $Fe_2SiO_4$, representing oxidized iron, is unstable at solar oxygen fugacities. This temperature is lower than the ca. 1400 K required for diffusion and annealing of silicate grains (Figs. 2 and 3). The solution is to pump up the oxidation state of the vapor phase by infusing it with water inside of the snowline. As the protoplanetary disk evolved, small icy bodies will have drifted inwards of the snowline, resulting in addition of $H_2O$ vapor by subsequent sublimation of water ice. The longevity of this vapor depended critically on the competing rates of inward migration across the snowline and the outward diffusive flux of water driven by freezing of water at the snowline (e.g., *Cuzzi and Zahnle, 2004*). The result of the competing fluxes may have been a pulse of elevated $H_2O/H_2$ just inside the snowline (Fig. 4). Such a localized pulse of water vapor has been observed in the TW Hydra transition disk at a distance of about 4 au from the central star (*Zhang et al., 2013*). The significant amount of oxidized iron in grains comprising chondrite matrix, even in rocks with limited evidence for water-rock reactions, is testament to the fact that such a process must have occurred where planetesimals were forming (*Grossmann et al., 2012*). The amount of water required can be estimated from the equilibrium relationship

$$H_2 + 1/2\, O_2 \Leftrightarrow H_2O$$

for which we obtain the equation for $f_{O_2}$ (e.g., *Krot et al., 2000*)

$$\log(f_{O_2}) \simeq 2\log\left(\frac{H_2O}{H_2}\right) + 5.67 - \frac{256664}{T(K)} \qquad (1)$$

Inspection of this equation shows that in order to change oxygen fugacity by ~ 5 log units as required to shift from solar to "planetary" oxidation states, $H_2O/H_2$ must have increased by a factor of ~ 375 at inner disk temperatures of ~1400 K (Fig. 5). Geometric considerations suggest that focusing solar abundances of water into the inner 5 au of the disk by inward transport can only raise the $H_2O/H_2$ ratio by about a factor of 10 (where, for example, surface density goes as $1/R$ where $R$ is the radial distance from the Sun).

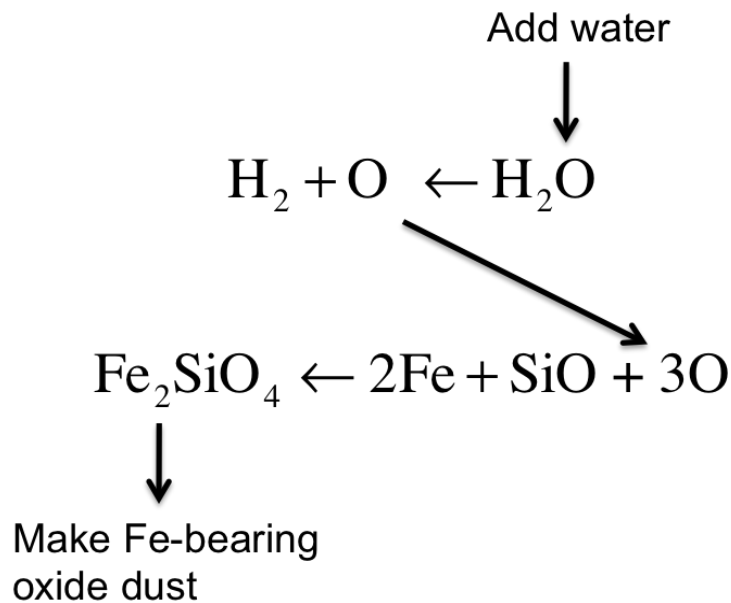

*Fig. 2: Schematic illustrating the enhancement of the fayalite component in olivine due to an increase in water. This scheme could have operated at moderately high temperatures in the protoplanetary disk where dust was processed.*

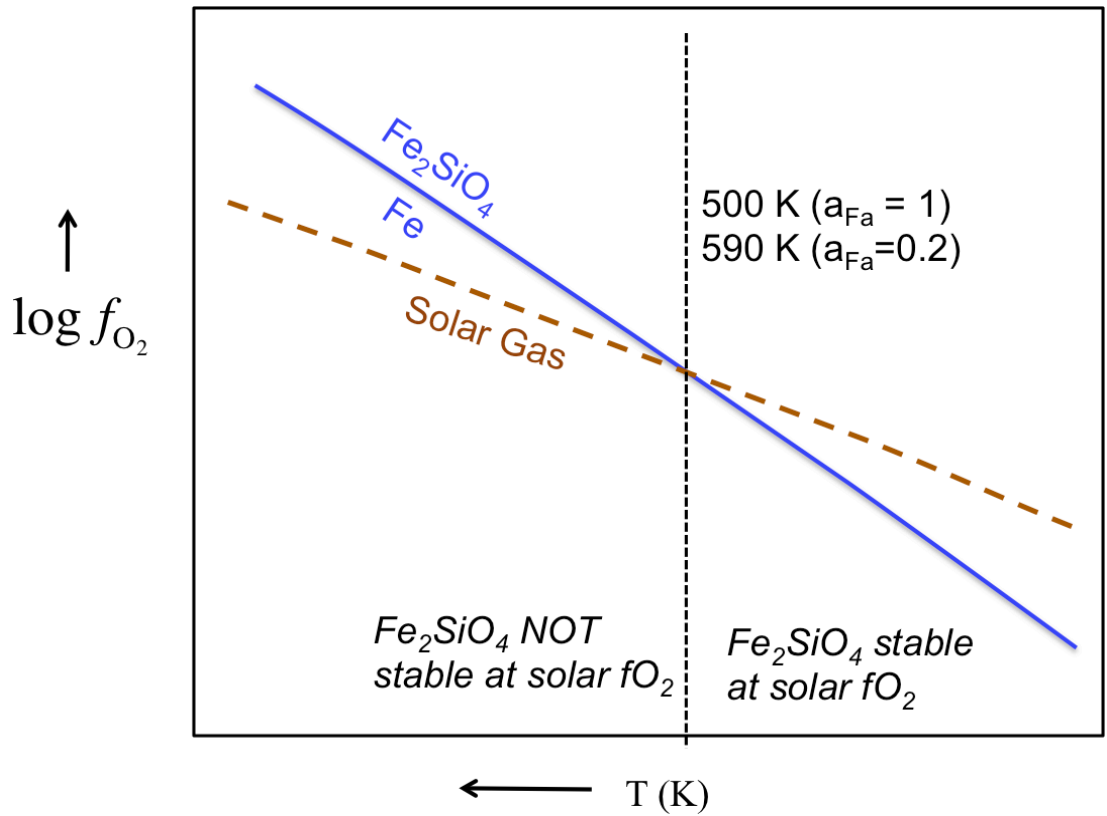

*Fig. 3: Schematic diagram showing the relationship between oxygen fugacity and temperature for a solar gas and for the oxidation state of iron. At higher temperatures, reduced iron is favored over the fayalite component for a solar-like gas. At lower temperatures, $Fe_2SiO_4$ can be stabilized, but whether this reaction can actually occur at lower temperatures is in question. Typical activities for fayalite in olivine at two different temperatures are shown for illustration.*

The picture that emerges is one in which ice accreted within planetesimals outside of the snowline may have reacted with the host rock as the planetesimals evolved in response to heating by $^{26}$Al (see also section 2). Liquid water reactions with rock will have oxidized the rock. Inside of the snowline there was likely a region where inward migration of ice enhanced the $H_2O/H_2$ ratio in the vapor phase just inside of the snowline, but further into the disk in the region of terrestrial planet formation, the ambient conditions favored reduced iron over oxidized iron as a result of higher temperatures and a decrease in gaseous $H_2O$ toward the star. The snowline must have migrated inward with time in response to a decrease in the stellar accretion rate, and the best fit shown in Fig. 1 suggests that its time-integrated "fossilized" location (see below) was somewhere between 6 and 2 au.

The position of the midplane water snowline as a function of time can be obtained from the temperature and pressure structure of a generic protoplanetary accretion disk. The $T$ and $P$ structure of the disk can be modeled using basic principles of dissipative radiative losses resulting from orbital torques. The torques are sustained by the viscosity of the disk that results in the inward spiral of gas towards the growing star. For dimensionless viscosity parameter $\alpha$ (e.g., *Shakura and Sumnyaev, 1973*), sound speed $c_s$, and scale height $H(R)$, the effective viscosity of the disk is $v = \alpha c_s H(R)$. The temperature at the midplane of the disk can be written in terms of the accretion rate of the system, $\dot{m}$, as follows

$$T_{\text{midplane}} = \left[\frac{\frac{3}{2}\Sigma(R)\kappa_R}{4} + 1\right]^{1/4} \left[\frac{3}{8\pi\sigma}\frac{GM_*\dot{m}}{R^3}\right]^{1/4} \quad (2)$$

where $\kappa_R$ is the Rosseland mean opacity, $\Sigma(R)$ is the surface density of the disk that is a function of $R$ from the central star, $G$ is the gravitational constant, $M_*$ is the mass of the central star, $\dot{m}$ is the accretion rate and $\sigma$ is the Stefan-Boltzmann constant. The first term on the right-hand-side of Equation (2) accounts for the optical depth perpendicular to the midplane and is equivalent to $T_{\text{midplane}}/T_{\text{surface}}$. The surface temperature is given by the second term. Equation (2) shows that the temperature of the disk at any radial distance $R$ from the star depends on the rate of mass accretion through the disk. The latter has been measured in young stellar objects today and varies from about $10^{-6}$ $M_\odot$ yr$^{-1}$ early in the history of stellar accretion to $10^{-8}$ $M_\odot$ yr$^{-1}$ (*Hartmann 2000*) later on.

Evaluation of Equation (2) requires evaluating the surface density with distance from the star, $\Sigma(R)$. The surface density can be written in terms of the viscosity as (*Armitage 2010*)

$$\Sigma = \frac{\dot{m}}{3\pi\alpha c_s H(R)}, \quad (3)$$

where scale height $H$ of the disk as a function of radial distance from the star is

$$H(R) = \frac{c_s}{\Omega} = \sqrt{\frac{k_b T R^3}{GM_*\mu}}, \quad (4)$$

and where $\Omega$ is orbital angular velocity, $k_b$ is Boltzmann's constant, $T$ is the temperature and $\mu$ is the average molecular mass for the gas comprising the disk. Equation (2) can be evaluated analytically with reasonable precision by assuming a constant value of 1 m$^2$/kg for $\kappa_R$ and replacing Equation (3) with a parameterization for $\Sigma(R)$ suggested by the mass distribution of planets in the solar system:

$\Sigma(R) \sim 1200$ kg/m² $(R/5\text{ au})^{-3/4}$ (compare with *Chiang and Youdin, 2010*). The result for $10^{-7}\ M_\odot\text{ yr}^{-1}$ is shown in Fig. 6.

The competing rates of temperature and pressure-dependent adsorption and desorption of water onto grain surfaces in the disk define the position of the water snow line. Pressure defines the total number density $n$ of molecules since $P = n_{\text{total}} k_b T$. Midplane pressures as a function of radial distance from the star are

$$P(R) = \frac{\Sigma}{\sqrt{\pi} H(R)} \frac{k_b T}{\mu}. \qquad (5)$$

The kinetic theory of gases yields the rate constant for adsorption of molecules onto grain surfaces:

$$k_{\text{ads}} = \langle \pi r_{\text{grains}}^2 \rangle \hat{V}_{\text{gas}} n_{\text{grains}} \qquad (6)$$

where $k_{\text{ads}}$ is the rate constant (s⁻¹), $r_{\text{grains}}$ is the average radius of the dust grains, $n_{\text{grains}}$ is the number density of the dust grains (cm⁻³), and $\hat{V}_{\text{gas}} = \sqrt{8 k_b T/(\pi \mu)}$ is the average gas velocity. *Hasegawa et al. (1992)* suggest a parameterization for the number density of grains based on total hydrogen such that $n_{\text{grains}} \sim 1.33 \times 10^{-12} n_H$ and $n_H = 2 \times n_{H2} \sim n_{\text{total}}$. Substitution of this expression into Equation (6), together with the mass of water molecules, yields in SI units

$$k_{\text{ads}} = \langle \pi r_{\text{grains}}^2 \rangle \sqrt{1.18 \times 10^4 T(K)}\, 1.33 \times 10^{-6} (P_{H_2}/(k_b T)) . \qquad (7)$$

Equation (7) can be evaluated using midplane temperatures and pressures (Equations 2, 4, and 5) and the fact that typical radii for dust grains relevant to the protoplanetary disk environment are 0.1 μm, or $1 \times 10^{-7}$ m. The competing rate constant for desorption of water from grains in SI units is *(Willacy et al., 1998)*:

$$k_{\text{des}} = 1 \times 10^{12} \exp(-4815/T(K)) \qquad (8)$$

At equilibrium rates of adsorption and desorption of water onto dust grains are equal so we have $k_{\text{ads}}\, n_{\text{H2O, gas}} = k_{\text{des}}\, n_{\text{H2O, grains}}$. Therefore, $k_{\text{ads}}/k_{\text{des}} \gg 1$ represents water frozen to grains and $k_{\text{ads}}/k_{\text{des}} \ll 1$ represents where most water molecules exist in the gas phase. A unit value for $k_{ads}/k_{des}$ defines conditions corresponding to the effective position of the snow line in the midplane of the disk. Substitution of Equations (2) and (5) into Equations (7) and (8) yields the plot in Fig. 7 showing the calculated position of the snowline in the disk as a function of accretion rate. The effective freeze out temperature for $\dot{m} = 10^{-8}\ M_{\odot}$/yr is ~190 K. The effective temperature for freeze out of water is 170 K for $\dot{m} = 10^{-7}\ M_{\odot}$/yr. Most importantly, we note that basic physics of accretion disks indicates that the calculated position of the snow line varies from ~ 5 au to 2 au when accretion rate varies from $10^{-7}$ $M_{\odot}$/yr to $10^{-8}\ M_{\odot}$/yr, respectively. This is the basis for the assertion that in a solar-like protoplanetary disk, we have a good idea of where the snowline should have been as a function of time.

Of course, radial drift complicates the simple picture described above. Morbidelli et al. (2016) point out that decoupling of radial drift velocities between dust and gas can result in a decoupling of "wet" and "dry" gas from the position of the snowline. The rates of icy particle drift inwards towards the star compared with the rates of planetesimal formation then become critical for understanding the compliments of water in protoplanetary materials, as described above in connection with Figure 1. Morbidelli et al. (2016) suggested that the apparent position of the water snowline recorded by planetesimal formation may be a "fossilized" snowline imprinted on the disk by the gap opened up by a growing Jupiter. Therefore, the rough correspondence between the position of the shift in oxygen fugacity in Figure 1 and the natural position of the water snowline for disk mass accretion rates of between $10^{-7}\ M_{\odot}$/yr and $10^{-8}\ M_{\odot}$/yr could be a manifestation of a locking in of the position of the snowline at these accretion rates by opening of a gap in the vicinity of the snowline by proto-Jupiter.

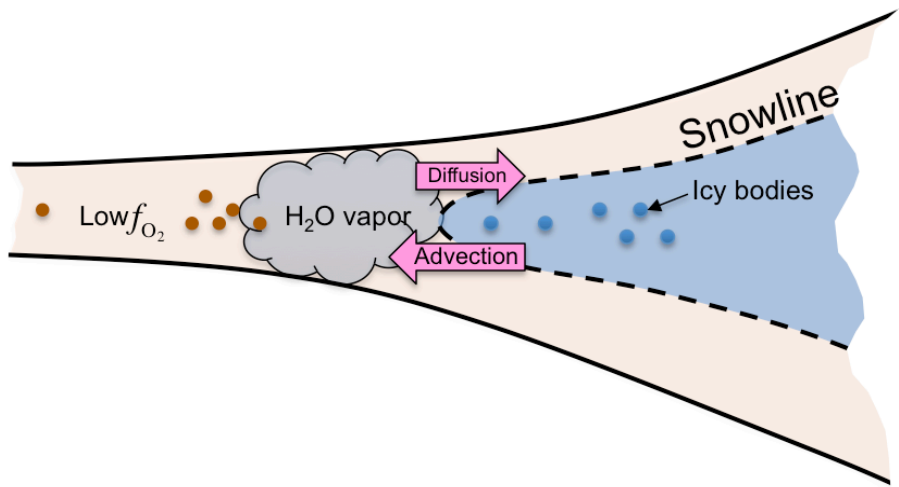

*Fig. 4: Schematic diagram illustrating the competition between advection of icy material through the snowline of the protoplanetary disk and backwards net transfer of water via diffusion driven by a buildup of water vapor inside of the snowline. The process is described by Cuzzi and Zahnle (2004) and is consistent with Spitzer and Herschel observations of TW Hydra (Zhang et al., 2013).*

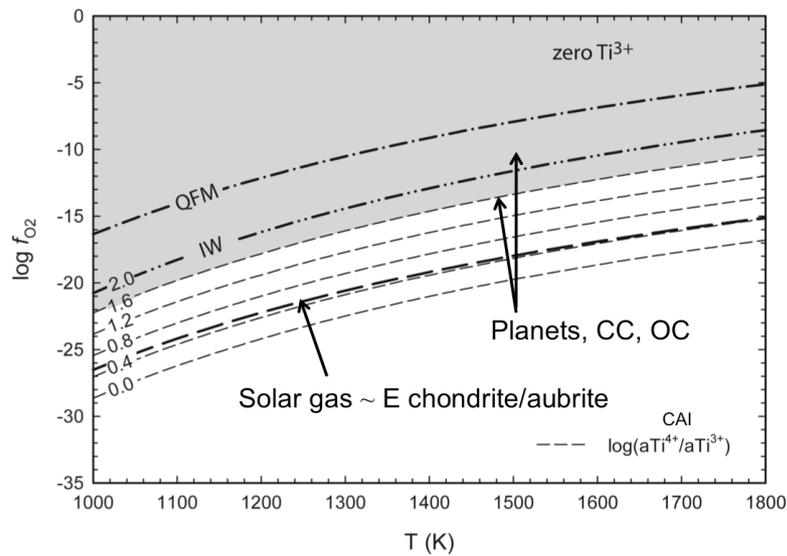

*Fig. 5: The logarithm of oxygen fugacity versus temperature for various oxygen buffers relevant for the early solar system. The heavy dashed line shows the relationship for a solar gas. For comparison, planetary materials (Earth's mantle, basalts from various bodies) exhibit oxygen fugacities closer to the iron-wüstite (IW) and quartz-fayalite-magnetite (QFM) oxygen buffers. Note that the difference between a solar gas and those planetary materials implies an increase of more than 5 orders of magnitude in oxygen fugacity. The region shaded in grey represents oxidation states that favor $Ti^{3+}$ over $Ti^{4+}$ in calcium-aluminum-rich inclusions (CAIs) in carbonaceous chondrites. Contours for the activity ratio of the two forms of Ti in CAIs are also shown.*

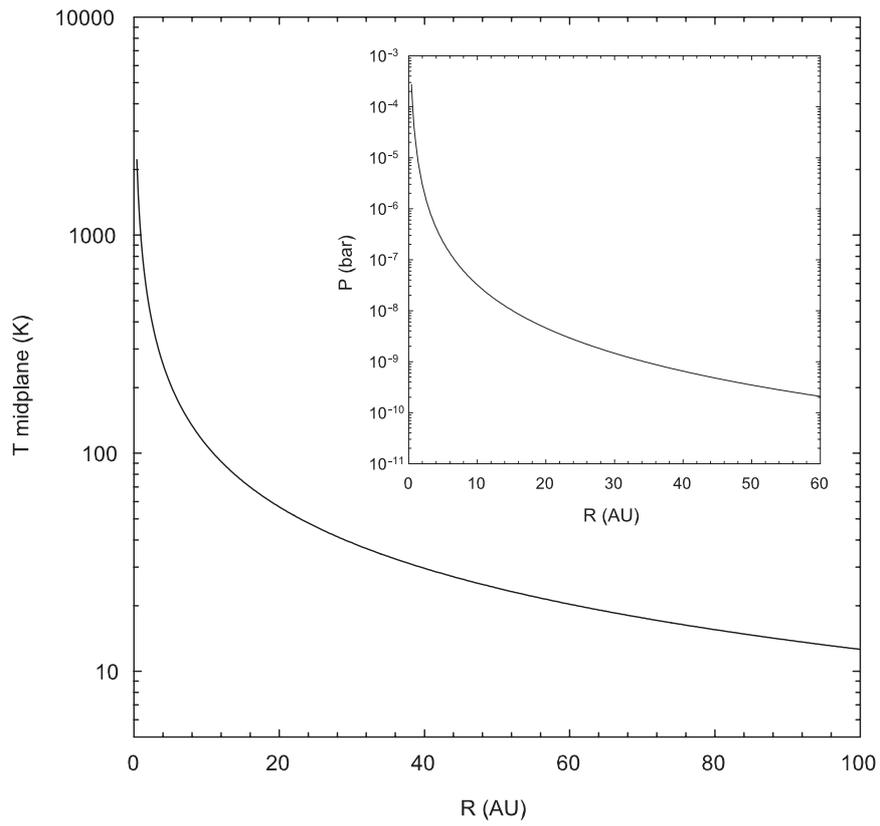

*Fig. 6: Calculated temperature and pressure vs. distance from the central star for a solar-like protoplanetary accretion disk with a mass accretion rate of $10^{-7}$ solar masses per year.*

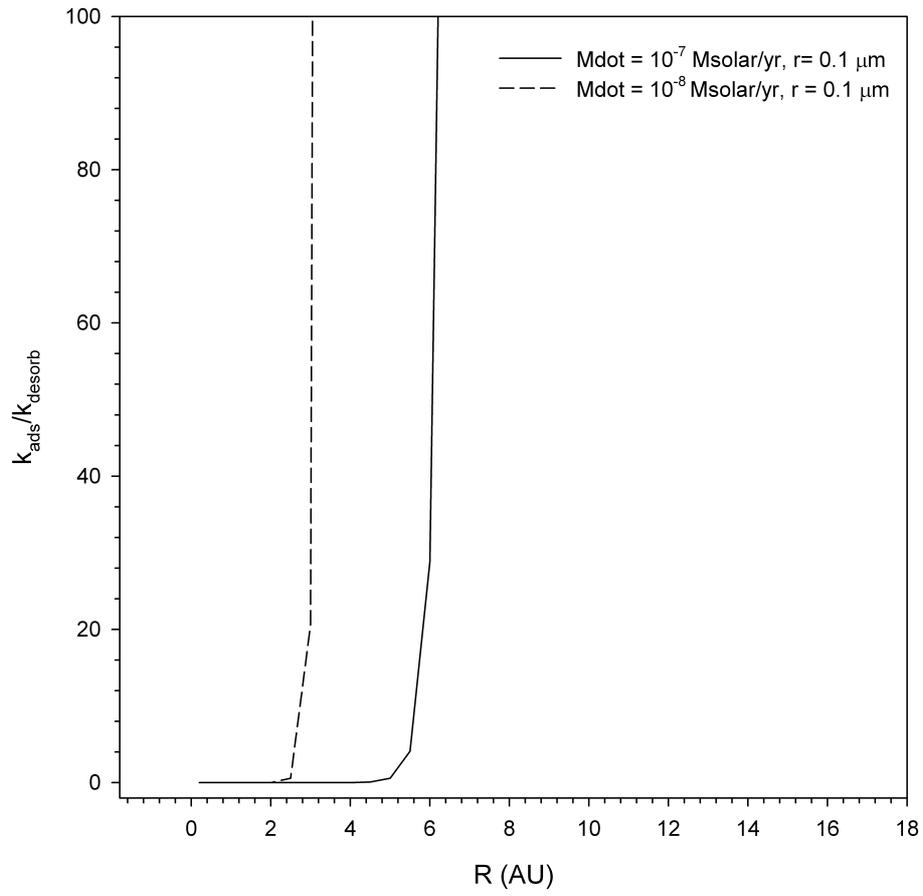

Fig. 7. Position of the water snow line as a function of mass accretion rate as defined by the ratio of the adsorption and desorption rate constants.

## 2. Effect of internal evolution on water

During the early evolution of the terrestrial planets and icy bodies, four main sources of energy may have contributed to their internal thermal budget (e.g. *Tobie et al. 2013*): (i) Tidal heating associated with despinning, (ii) radiogenic heating by the decay of short-lived radioactive isotopes, (iii) impact heating and (iv) conversion of potential energy into heat via viscous dissipation during core-mantle separation. These four sources of energy likely concern different growth stages or reservoirs and hence will be separated in time and space. For planetesimals, only short-lived radiogenic (mainly $^{26}$Al) heating is relevant (*Yoshino et al., 2003; Rubie et al., 2015b*). This source of energy was for several million times larger than the radiogenic heating rate due to the decay of long-lived isotopes currently expected on planets and moons. However, this huge heat production decayed rapidly. On the other hand on planetary embryos and fully accreted terrestrial planets, the release of potential energy and impact heating have to be taken into account in addition to radiogenic heating. The impact heating strongly depends on the impact velocity, on the masses of the target and the impactor as well as the way kinetic energy is converted into heat (*Monteux et al., 2007*). Hence the size of the growing body that governs the minimum impact velocity (i.e. the escape velocity) is a key factor. For large growing bodies, large impact events may generate significant melting within the target and trigger metal-rock or ice-rock separation potentially leading to runaway differentiation processes (*Golabek et al., 2009; Ricard et al., 2009*). Finally on icy moons long-lived radiogenic heating and tidal dissipation are the most important energy sources, while only the largest impact heating events can play a role (*Monteux et al., 2014*).

The composition of the building blocks from which planetary objects form as well as the early thermal budget should control the amount of water that can be stored in planetary embryos. To get a first order estimation of the amount of hydrous silicates and leftover rock-ice mixture available at the end of the thermo-mechanical evolution of planetesimals, we perform numerical models in 2D infinite cylinder geometry using the finite-difference marker-in-cell code I2ELVIS (*Gerya and Yuen, 2007; Golabek et al. 2014; Lichtenberg et al., 2016*). The physical parameters and the setup employed are identical to those used in our previous studies (see *Golabek et al. 2014; Lichtenberg et al., 2016*).

Based on current planetesimal formation models (e.g. *Johansen et al., 2015; Simon et al., 2016*), we model planetesimals with radii $R_P$ ranging between 25 and 230 km and consider instantaneous formation times $t_{form}$ ranging from 0 to 3.5 Myr after formation of Ca-Al-rich inclusions (CAIs). We study here the evolution of planetesimals that formed across the snowline and were scattered later into the inner solar system (*Raymond and Izidoro, 2017*). Thus we assume for each planetesimal a starting temperature equal to the temperature of surrounding space: $T_{space}$ = 150 K. Since for small objects the release of potential energy related to accretion is small (*Schubert et al., 1986*), the resulting temperature increase is negligible and is not considered here. Thus only short-lived radiogenic heating is relevant. For simplicity we assume that the planetesimals do not contain metal because both formation of hydrated silicates and their dehydration occur at lower temperatures than the melting temperature of the eutectic Fe-FeS mixture (1243 K) at low pressures (*Brett and Bell, 1969*), so core formation due to percolation is expected to start only after the breakdown of hydrated silicates. Therefore we assume that at the start of the model each planetesimal is composed only of a rock-ice mixture. Here we assume that at $T = 273$ K the ice melts and hydrated silicates can form. Under the low-pressure conditions inside a planetesimal the most temperature-resistant hydrated silicates (amphibolites) break down at $T \approx 1223$ K and dehydrate (*Fu and Elkins-Tanton, 2014*). However it should be taken into account that by considering the breakdown of amphibolites the current models give only an upper limit for the amount of hydrated silicates present, since for example more abundant hydrated silicates like serpentine-phyllosilicates break down at significantly lower temperatures (573-673 K) (*Nakamura, 2006*; *Nakato et al., 2008*).

The following additional assumptions and simplifications are used in the numerical model: We ignore here reaction kinetics and assume that both the formation of hydrated silicates and their breakdown occur instantaneously at the given temperatures. Also we ignore possible water delivery or loss processes related to impact events. Additionally we do not consider pore water convection (*Young et al., 2003*) suggested for smaller planetesimals ($R_P < 40\text{-}60$ km) and assume that after silicate dehydration the water previously incorporated into hydrated silicates is completely lost to space. Also it should be kept in mind that even dry silicates still contain small amounts of water (see e.g. *Fu and Elkins-Tanton, 2014*), thus even planetesimals loosing all hydrated silicates still contain some water, which is not considered in our numerical model.

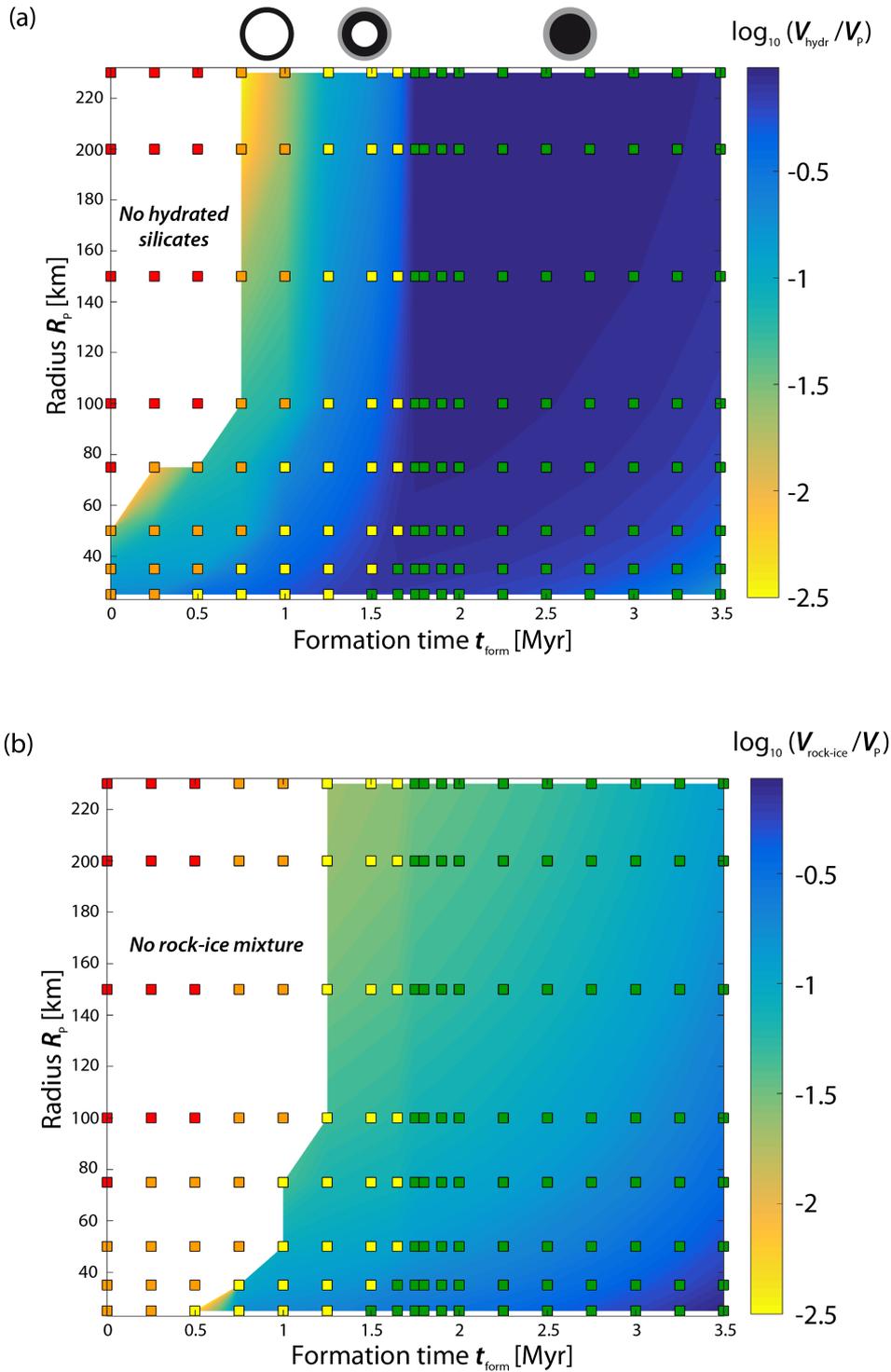

*Fig 8: (a) Final volume of hydrated silicates $V_{hydr}$ and (b) Final volume of rock-ice mixture $V_{rock\text{-}ice}$. Both volumes are scaled by the total volume of the specific planetesimal $V_P$ for models with different planetesimal radii $R_P$ and formation times $t_{form}$ assuming an initial temperature $T_{space} = 150$ K. Squares represent the numerical model results. Colours of the squares stand for various final internal states: No hydrous silicates left (red), outer hydrated shell (orange), inner hydrated shell (yellow) and hydrated interior (green). For clarity three of the possible final outcomes are sketched on top of subfigure (a) with dehydrated silicates (white), hydrated silicates (black) and rock-ice mixture (grey).*

The time-dependent amount of hydrated silicate material in each modeled planetesimal is tracked on the markers of the numerical model. The final 2D area covered by (a) rock-ice mixture and (b) hydrated silicates is converted for representation purposes analytically into a 3D volume (see Fig. 8 a+b).

The general results show that the amount of both remaining rock-ice mixture and hydrated silicates is strongly dependent on the formation time $t_{start}$ since this parameter controls the abundance of short-lived $^{26}$Al ($t_{1/2}$ = 0.716 Myr), while the radius has only a modest effect (see Fig. 8 a+b). As expected late formed objects contain orders of magnitude more rock-ice mixture and hydrated silicates than early-formed planetesimals. Since formation of hydrated silicates requires some heating, more of the interior contains hydrated silicates in late-formed, but large planetesimals, while a primordial rock-ice mixture is best preserved in the smallest objects experiencing fast cooling. However the results also show that less hydrated silicates form in planetesimals that accreted very late because the reduced amount of $^{26}$Al radiogenic heating is insufficient to allow for the formation of hydrated silicates throughout the bulk of the planetesimal interior. This leads to a "sweet spot" for the formation of hydrated silicates, which is narrower for small planetesimals with $R_P$ < 80 km since these objects cool more efficiently than large planetesimals (see Fig. 8a).

All models start with a rock-ice mixture. Because the temperature at the center of the planetesimal reaches the maximum value, formation of hydrated silicates starts there and over time the formation front propagates outwards. For large objects ($R_P \geq$ 100 km) that formed early enough ($t_{form}$ < 0.75 Myr) the center is also the location where dehydration starts during later evolution and a dehydration front propagates from there towards the surface. For planetesimals with sufficient $^{26}$Al heating both the formation and later the dehydration front are able to propagate throughout the entire planetesimal interior and within ~$10^6$ years the entire planetesimal will loose all hydrated silicates (see Fig. 9). On smaller objects full dehydration occurs only on objects that formed earlier than $t_{form}$ < 0.75 Myr, while in the smallest planetesimals considered ($R_P \leq$ 50 km) complete loss of hydrated silicates is not achieved even inside planetesimals formed at the same time as CAIs (see Fig. 8a).

Large, but later-formed planetesimals (0.75 Ma $\leq t_{form}$ < 1.25 Myr) or early-formed, but small objects exhibit a different final outcome, namely an outer hydrated shell, where the formation front reached the object's surface, while the propagation of the dehydration front ceased at depth due to insufficient $^{26}$Al radiogenic heating. Inside large planetesimals that formed even later (1.25 Ma $\leq t_{form}$ < 1.75 Myr) neither the formation nor the dehydration front reach the surface. Under these circumstances the surface is still composed of a rock-ice mixture while the deep interior is dehydrated and an annulus at intermediate depth contains hydrated silicates. For late-formed objects (large objects with $t_{form} \geq$ 1.75 Myr or for smaller planetesimals with $t_{form}$ ~ 1.5 Myr after CAI formation) temperatures are never sufficiently high as to start dehydration and these objects display a hydrated interior and a rock-ice mixture closer to the surface (see Fig. 8 a+b). This is in agreement with observations indicating that some asteroids contain even at present-day significant amounts of volatiles (*Hsieh and Jewitt, 2006*; *Jewitt et al., 2009*; *Küppers et al., 2014*).

Based on these simple models, we can speculate that large ($R_P > 80$ km), late-formed planetesimals ($t_{form} > 1.7$ Myr), initially formed across the snowline and implanted into the main asteroid belt or scattered into the inner solar system during the accretion of the gas giants (*Raymond and Izidoro, 2017*), can contribute significant amounts of water in the form of hydrated silicates to growing planetary embryos. This is in general agreement with both the classical model (e.g. *Chambers, 2013*; *Raymond et al., 2009*) and the Grand Tack scenario (*Walsh et al., 2011*; *Rubie et al., 2015a*) suggesting that during the later stages of accretion planetary embryos will accrete material that formed at larger distances from the Sun.

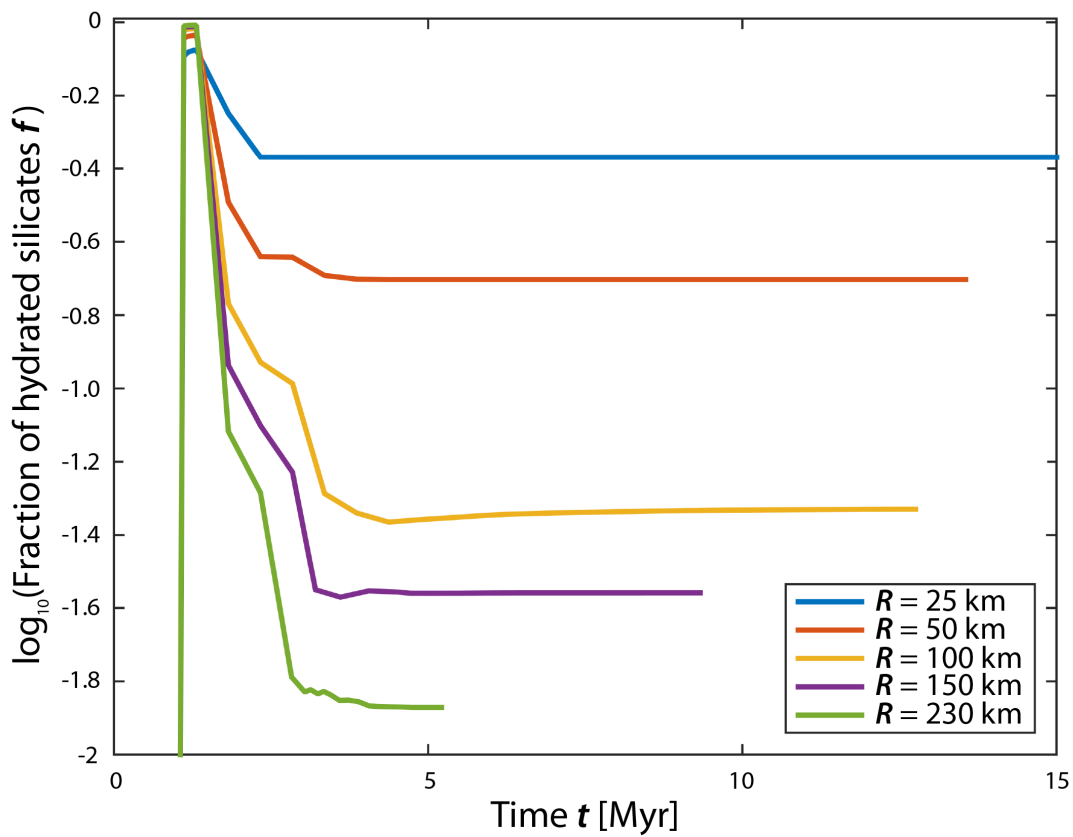

*Fig. 9: Time-dependent fraction of hydrated silicates inside planetesimals with $t_{form} = 1$ Myr and $R_P = 25$-$230$ km.*

In conclusion, we have shown that the internal evolution and especially the formation time of planetesimals relative to the timescale of radiogenic heating by short-lived $^{26}$Al decay may govern the amount of hydrous silicates and leftover rock-ice mixture available during the late stages of their evolution. In turn, water content may affect the early internal evolution of the planetesimals and in particular the metal-silicate separation processes to be discussed in the next section.

## 3. Effects of water on internal evolution

The most important process occurring during the early internal evolution of a rocky planetary body is differentiation that results in a silicate mantle and a metallic iron-rich core. The terrestrial planets, Mercury, Venus, Earth and Mars, as well some asteroids, such as Vesta, all underwent core-mantle differentiation. In addition, based on the existence of iron meteorites, many early-formed planetesimals also experienced core-mantle differentiation. Undifferentiated material, as represented by chondritic meteorites, consists of intimate mixtures of metal, sulfide, silicate and oxide grains. The formation of metallic cores in differentiated bodies therefore requires metal and silicate to separate over significant length scales (up to 3000 km in the case of Earth but on the order of 100-500 km in the case of planetesimals). Such segregation was only physically possible at high temperatures at which at least the metal and probably also the silicates were in a molten state (*Stevenson, 1990; Rubie et al., 2015b*). High temperatures were the result of the decay of the short-lived $^{26}$Al isotope in planetesimals during the first ~3 Myr of Solar System history and were later the result of high-energy collisions that took place during planetary accretion (e.g. *Rubie et al., 2015b; de Vries et al., 2016*). Thus planetesimals could only differentiate early, while $^{26}$Al was extant, and were largely unaffected by impact-induced heating because of their low masses (see *Rubie et al., 2015b*, Fig. 10). In general, core-mantle differentiation that occurred as a consequence of impact-induced melting could only happen later after bodies had grown sufficiently in mass. Physical and chemical conditions differed significantly during these respective episodes. For example, gravitational forces were much smaller during early planetesimal differentiation and such bodies likely differentiated in the presence of nebular gases, which dispersed before the formation of the terrestrial planets.

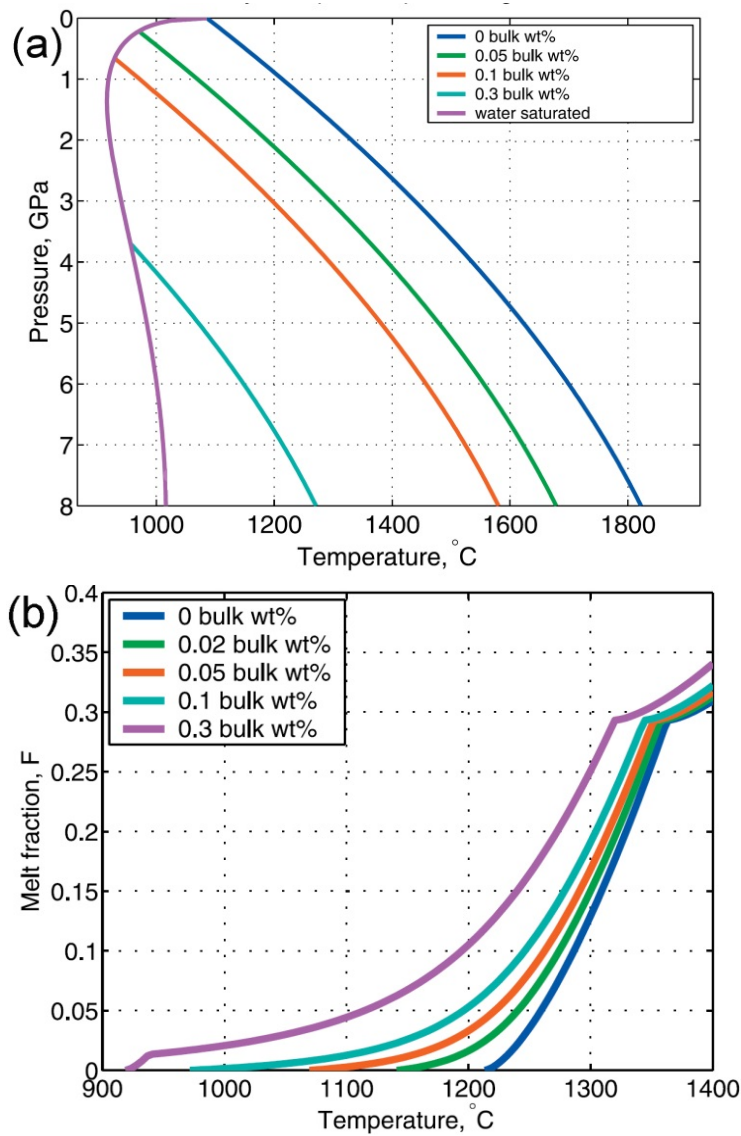

*Fig. 10. Effects of water on the melting of peridotite. (a) Depression of the peridotite solidus temperature as a function of pressure for bulk water contents of 0, 0.05, 0.1 and 0.3 wt%. (b) Effect of water on melt fraction F during partial melting of peridotite shown as isobaric melting curves at 1 GPa for bulk water contents ranging from 0 to 0.3 wt%. Note that the kink at a melt fraction of ~0.3 is caused by the disappearance of clinopyroxene from the mineral assemblage (From Katz et al., 2003). Note that equivalent temperatures for a bulk chondritic composition are lower by ~200 °C (Fig. 7 in Asahara et al., 2004). Also the addition of any incompatible elements (e.g. the C and S that are present in chondritic meteorites) will reduce melting temperatures.*

The presence of water in planetary bodies could have had a significant effect on both the physical processes of core formation and their chemical consequences. For example, the presence of even small amounts of water strongly reduces the solidus temperature of silicates (Fig. 10a). It also increases the melt fraction at a given temperature, although this effect is largest close to the solidus and becomes much smaller as the degree of melting increases (Fig. 10b).

*2.1 Effects of water on mechanisms of core formation*

There are several different mechanisms by which metal and silicate could have segregated (Fig. 11) and the presence of water would have affected each of these differently. Potentially the mechanisms shown in Fig. 11 could all operate during the differentiation of an Earth-mass planet. However, as discussed towards the end of this section, core-mantle differentiation on small bodies (planetesimals) likely required significant melting (with melt fraction of at least 30-50%) as the result of heat produced by the decay of $^{26}$Al (see also *Scheinberg et al., 2015*).

We first review the possible metal-silicate segregation mechanisms and how these would be affected by the presence of water and then discuss chemical consequences of water during core-mantle differentiation.

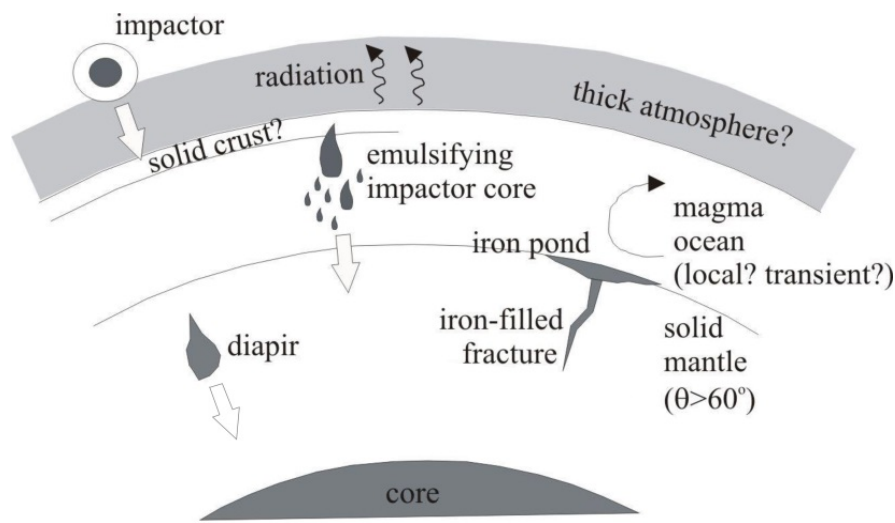

*Fig. 11. Schematic section through a terrestrial planet during core formation. A global magma ocean is shown overlying solid crystalline mantle below which lies the metallic proto-core. Metal is delivered by impactors, many of which are likely to have been differentiated into core and mantle. The molten impactor cores partially or completely emulsify in the magma ocean to form small droplets that sink to the base of the magma ocean and collect there to form liquid "iron ponds". These sink through the crystalline mantle as diapirs or possibly are transported through fractures (see Rubie et al., 2015b for a review). Although a possible solid crust is shown, this may not form when an atmosphere is present because its insulating effect keeps the temperature at the surface of a magma ocean above the solidus (e.g., Matsui and Abe, 1986).*

<u>*Percolation of liquid metal through a polycrystalline or partially-molten silicate matrix.*</u> Liquid iron-rich metal or sulfide, (Fe,Ni)S, can percolate through a solid polycrystalline silicate matrix by porous flow provided the liquid phase forms an interconnected network along grain boundaries or grain edges rather than isolated melt pockets. Interconnection depends upon the dihedral angle, $\theta$, which is the angle between two solid–liquid boundaries that are intersected by a solid–solid boundary at a triple junction (Fig. 12; *von Bargen and Waff, 1986; Stevenson, 1990; Rubie et al., 2015b*). When $\theta$ is less than 60°, the liquid metal phase is interconnected at any melt fraction and can percolate efficiently through the crystalline matrix. When $\theta$ exceeds 60°, interconnection only occurs when a critical melt fraction (e.g. a few volume % or more) is exceeded, which means that metal-silicate segregation cannot occur efficiently. Experimental studies of both natural and synthetic systems relevant to core formation have shown that, in general, $\theta$ lies in the range 80-120°, at least up to pressures of ~25 GPa (*Rubie et al., 2015b and references therein*). Therefore percolation is unlikely to be a major mechanism of metal-silicate segregation in the terrestrial planets. However, at pressures <2-3 GPa, the dihedral angle is less than 60° if the liquid Fe contains significant concentrations of oxygen and sulfur (*Terasaki et al., 2005, 2008*). It is therefore possible that under oxidizing conditions percolation was an important core formation mechanism in planetesimals during heating by $^{26}$Al decay, especially when the metal contains sulfur (which reduces the melting temperature of the metal relative to that of the silicate).

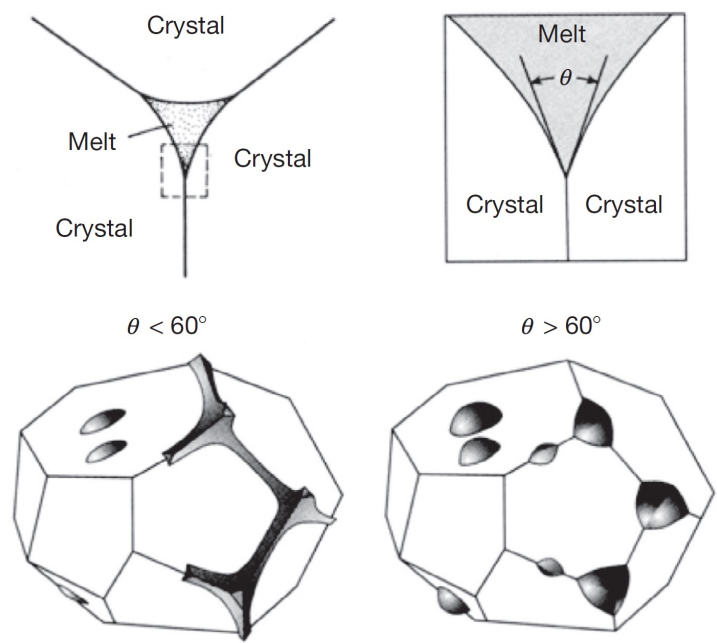

*Fig. 12. The dihedral angle $\theta$ is the angle between two melt-crystal interfaces where they are intersected by a crystal-crystal grain boundary at a triple junction. When $\theta$<60°, the melt forms an interconnected network along grain edges (bottom left), whereas when $\theta$>60° there is no interconnectivity (unless the melt fraction is high) (from Stevenson, 1990).*

As discussed in the previous section, dehydration of silicates is likely to leave only a few hundred ppm of water in planetesimals that have been heated by the decay of $^{26}$Al. However, such H$_2$O concentrations would be sufficient to lower the silicate solidus temperature by 100-150 K (Fig. 10a), which means that the percolation of liquid metal or sulfide through partially molten silicates, rather than crystalline silicates, needs to be considered. Intuitively, it might be expected that a few percent silicate liquid in the system would enhance the ability of liquid metal to percolate. However, the opposite is the case and the presence of silicate melt actually reduces the interconnectivity of metallic liquids (*Holzheid et al., 2000; Rushmer and Petford, 2011; Holzheid, 2013; Cerantola et al., 2015; Todd et al., 2016*). This is the result of the high liquid metal-liquid silicate interfacial energy that causes metallic blobs to adopt a spherical geometry so that they become isolated in silicate melt. Consequently the metallic blobs become trapped in interstices in the partial molten aggregate. Thus, once silicate liquid forms by partial melting, as temperature increases, any interconnectivity of liquid metal or sulfide is destroyed. As the temperature rises further, the silicate melt fraction will eventually become large enough to enable the metallic droplets to become mobile and to segregate gravitationally. The value of the silicate melt fraction at which this happens is poorly known but is likely to be in the range 30-50% (*Stevenson, 1990; Minarik et al., 1996; Holzheid et al., 2000; Costa et al., 2009; Solomatov, 2015*). At this degree of partial melting a magma ocean forms and the segregation mechanism changes to that described below. Thus the presence of water is likely to significantly delay metal-silicate segregation in planetesimals by requiring a higher temperature to start than under anhydrous conditions.

*Segregation of liquid metal in a silicate magma ocean.* As described above, when the fraction of silicate melt exceeds a certain value (probably 30-50%), a magma ocean state is achieved. When the metallic cores of impacting bodies sink through a magma ocean by gravitational settling, they tend to break up into smaller masses, eventually forming small droplets through emulsification. The extent to which emulsification occurs is important because it determines the fraction of accreted metal that equilibrates chemically with silicate liquid of the magma ocean (e.g., *Nimmo et al., 2010; Rudge et al., 2010; Rubie et al., 2015a*). When emulsification and equilibration fail to occur, the impactor's core merges directly with the core of the target body by "core merging". The efficiency of emulsification is uncertain, especially for the cores of giant impactors (*Rubie et al., 2003; Dahl and Stevenson, 2010; Samuel, 2012; Deguen et al., 2014; Wacheul et al., 2014; Kendall and Melosh 2016; Landeau et al., 2016*). The process is extremely difficult to study numerically because the length scales involved range from 100's km (or more) to a few mm. In the case of the Earth, it has been estimated that emulsified droplets of liquid Fe have a stable diameter of ~1 cm and sink with a velocity of ~0.5 m/s (*Stevenson, 1990; Rubie et al., 2003*). Such parameters are a result of molten peridotite having a very low viscosity (*Liebske et al., 2005*) – which also results in magma oceans convecting vigorously in a turbulent regime. Because peridotite liquid is already highly depolymerized, the addition of water is unlikely to have a significant effect on its viscosity. In addition, as shown in Fig. 10b, the presence of water will have only a small effect on the temperature of magma ocean formation when the latter is defined to have a silicate melt fraction $\geq 0.3$.

*Descent of diapirs of liquid metal through crystalline mantle.* It has been proposed that segregated liquid metal accumulates as a pool at the bottom of a magma ocean during the differentiation of terrestrial planets. If the base of the magma ocean is separated from the proto-core by a crystalline mantle, it is likely that the iron melt pool sinks through the latter as diapirs, perhaps 1-10 km in diameter or larger (Fig. 11; e.g. *Stevenson, 1990; Karato and Murphy, 1997; Samuel et al., 2010*). Water dissolved in crystalline upper mantle rocks has a strong weakening effect on rheology and reduces the viscosity of solid-state flow by up to 2-3 orders of magnitude. For example, *Hirth and Kohlstedt (1996)* show that 810 ± 490 H/$10^6$ Si dissolved in olivine reduces the viscosity by a factor of 500 ± 300 compared with that of dry olivine. The descent velocity of metal diapirs can be described by Stokes' Law. Since the Stokes' sinking velocity is inversely proportional to mantle viscosity, the effect of water is large and could enhance the descent velocity by up to 2-3 orders of magnitude, at least in upper mantle peridotite.

Numerical models of diapirs sinking through the solid mantle of Earth-sized bodies indicate that a wet rheology favors diapir sinking, while a dry rheology leads to preferential formation of iron-filled dikes (*Golabek et al. 2009*). It was shown both analytically and numerically that this changes the heat partitioning between iron and silicates (*Ke & Solomatov, 2009; Golabek et al. 2009*). Whereas the sinking of diapirs results in heat preferentially partitioning into the silicate mantle, the propagation of iron dikes results in the preferential heating of the iron. Thus the presence of water in the crystalline mantle could affect the thermal state of a terrestrial planet after core formation.

In summary, the effect of water on the temperature and timing of magma ocean formation in accreting planetary bodies, as required for metal-silicate segregation, is very small. In relatively oxidized planetesimals, the presence of water may eliminate the possibility of metal-silicate segregation by percolation, as proposed by *Terasaki et al. (2008),* so that segregation only starts at a magma ocean stage after mantle material is at least 30-50% molten. If the descent of diapirs of liquid metal through crystalline mantle occurs, water dissolved in the mantle material would greatly enhance the rate of this process. On the other hand, an absence of water could change the descent mode from sinking diapirs to diking, thus affecting the heat partitioning between iron and silicates and modifying the thermal state of the early terrestrial planet.

*2.2 Effects of water on the chemistry of core formation*

The main geochemical effect of metal-silicate segregation to form the core and mantle of a planetary body is that siderophile (metal-loving) elements, such as Ni, Co, V, Cr, W, Mo, etc., partition into liquid metal and are thus transported from the mantle into the core. Consequently, all siderophile elements are variably depleted in Earth's mantle, relative to solar relative abundances, by factors of up to almost 1000. The degree of depletion depends on the extent to which an element partitions into metal. This is described by the metal-silicate partition coefficient, $D_M^{met-sil}$, which for element M is defined simply as:

$$D_M^{met-sil} = \frac{C_M^{met}}{C_M^{sil}}$$

where $C_M^{met}$ and $C_M^{sil}$ are the molar (or wt%) concentrations of M in the metal and silicate, respectively. Most metal-silicate partition coefficients for siderophile elements depend on pressure (*P*) and/or temperature (*T*). In addition, and of great importance for the present discussion, they depend strongly on oxygen fugacity $f_{O2}$, as shown by the redox reaction:

$$M + (n/4)O_2 = MO_{n/2}$$
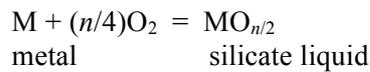
metal             silicate liquid

where *n* is the valence of element M when incorporated in silicate liquid as an oxide component. When $f_{O2}$ is low, this reaction shifts to the left thus increasing $D_M^{met-sil}$, whereas when $f_{O2}$ is high the reaction shifts to the right thus decreasing $D_M^{met-sil}$.

When modeling core formation, the extent to which Earth's mantle has been depleted in siderophile elements enables estimates to be made of the conditions (*P*, *T*, $f_{O2}$) of metal-silicate equilibration during core formation (e.g. *Wade and Wood, 2005; Walter and Cottrell, 2013; Rubie et al., 2011, 2015a, 2016*). Traditionally, it has been assumed that core formation was a single-stage event, thus enabling unique values for *P*, *T* and $f_{O2}$ to be determined (*Walter and Cottrell, 2013*; see Table 3 in *Rubie et al., 2015b*). However, "single-stage" core formation is clearly unrealistic because metal is delivered to a growing planet during the entire accretion process, which in the case of Earth likely lasted for up to 100 Myr (e.g. *Jacobson et al., 2014*). Thus, in recent years more realistic core formation models have been formulated. Firstly, in models of "continuous" core formation (*Wade and Wood, 2005; Wood et al., 2006*), material is accreted to Earth in small (1%) increments with the metal component being equilibrated with the silicate mantle at the base of a magma ocean. In a second class of models, core formation is multi-stage and the accretion of mostly-differentiated embryos and planetesimals to growing planets is linked to episodes of metal-silicate equilibration and core formation (*Rubie et al., 2011, 2015a, 2016*). In both types of models, magma ocean depth increases as planets grow so that the pressure and temperature of metal-silicate equilibration both increase as accretion proceeds. In addition, the oxygen fugacity at which metal and silicate equilibrate has to increase by 2-3 orders of magnitude during accretion in order to reproduce the concentrations of siderophile elements in Earth's mantle (*Wade and Wood, 2005; Wood et al., 2006; Rubie et al., 2011*). There are at least three mechanisms that can result in, or contribute to, this increase of $f_{O2}$, two of which are intimately related to water. Note that oxygen fugacity is determined from the compositions of co-existing metal and silicate relative to that of the iron-wüstite (IW) buffer, as follows:

$$\Delta \text{IW} = 2 \log \left[ \frac{a_{\text{FeO}}}{a_{\text{Fe}}} \right].$$

Here $\Delta$IW is the deviation of $f_{O2}$ from the IW buffer in log units, and $a_{\text{FeO}}$ and $a_{\text{Fe}}$ are the activities of FeO and Fe in the silicate and metal, respectively. Under low oxygen fugacities (e.g. $\Delta$IW = -4) the FeO content of the silicate is low (e.g. <1 wt%), whereas at higher oxygen fugacities (e.g. $\Delta$IW = -2) the FeO content becomes similar to that of Earth's mantle (~8 wt%)

*Partitioning of silicon into the core.* Under reducing conditions and especially at high temperature, Si becomes siderophile (e.g. *Mann et al., 2009*) and this element partitions into liquid core-forming metal by the reaction

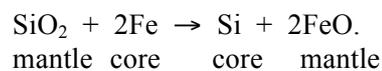

$$\begin{array}{cccc} \text{SiO}_2 + & 2\text{Fe} & \rightarrow \text{Si} + & 2\text{FeO}. \\ \text{mantle} & \text{core} & \text{core} & \text{mantle} \end{array}$$

According to this reaction, for every mole of Si that partitions into the core, two moles of FeO are added to the mantle. This reaction is therefore very efficient at increasing the oxygen fugacity, as required by the core formation models of *Wade and Wood (2005)* and *Rubie et al. (2011, 2015b)*.

*Oxidation of metal by accreted water.* At the end of core formation, the oxygen fugacity of Earth's mantle was low (~IW-2) whereas it is currently several orders of magnitude higher (close to the FMQ buffer). In the upper mantle, the currently high oxygen fugacity is the result of a high $Fe^{3+}$ concentration in some of the constituent minerals (*O'Neill et al., 1993*). The increase in $f_{O2}$ after core formation may have been caused by the disproportionation reaction

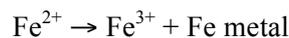

$$Fe^{2+} \rightarrow Fe^{3+} + \text{Fe metal}$$

which occurs as a result of the crystallization of the lower-mantle mineral bridgmanite (silicate perovskite) (*Frost et al., 2004*). Water can also potentially cause oxidization in planetary interiors, for example by the reaction:

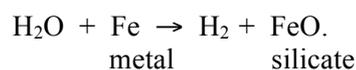

$$\begin{array}{cccc} \text{H}_2\text{O} + & \text{Fe} & \rightarrow \text{H}_2 + & \text{FeO}. \\ & \text{metal} & & \text{silicate} \end{array}$$

for which the loss of hydrogen is required, or, in the absence of metal, by the dissociation of $H_2O$ into oxygen and hydrogen, with the latter being lost. *Sharp et al. (2013)* proposed that such reactions were the cause of the increase in $f_{O2}$ of Earth's mantle following core formation.

All oxidation reactions that are caused by water require that hydrogen is lost from the system. In the absence of metal, this may be problematic because hydrogen is a strong reducing agent. Therefore, if oxidation is taking place deep in the Earth's interior, the released hydrogen has to migrate to the

Earth's surface without causing any reduction of the material that it comes into contact with, which seems very unlikely. If metal is present, the hydrogen may be removed by dissolving in the metal, so that, during core formation, it would be transported to the core of the planet. It is uncertain how efficient this process would be because the partitioning of hydrogen between metal and silicate under core formation conditions is poorly known (*Okuchi, 1997*). The oxidation of metal by water during core formation has been included in the accretion/core formation model of *Rubie et al. (2015b)* assuming that all released hydrogen dissolves in the metal. It was thus shown that the role of accreted water in oxidizing the mantle during core formation is insignificant and that all oxidation (up to the end of core formation) can be explained by the dissolution of Si into the core (as discussed above) combined with the effects of the accretion of oxidized bodies due to an oxidation gradient in the solar nebula (Fig. 1).

Most of the physical processes that are discussed to occur during the early stages of planetary formation also occur in the outer part of the Solar System during the formation of icy moons. In these bodies the mantle is composed mainly of water ice while their core is essentially made of rocky material. While the variety of internal structures among the icy moons is still debated, the essential role of the water is widely accepted. These objects will be discussed in the next section.

## 4. Differentiation and evolution of water-rich objects: from icy moons to ocean-planets

*4.1 Variety of internal structure of icy moons:*

Differences in composition and internal structure exist among the major icy satellites of Jupiter and Saturn, suggesting distinct accretion and differentiation histories (e.g., *Kirk and Stevenson, 1987; Mueller and McKinnon, 1988; Mosqueira and Estrada, 2003; Barr and Canup, 2008; Monteux et al, 2014*). In this section we will focus particularly on the largest icy moons having a planetary scale, namely Jupiter's moons Ganymede and Callisto, and Saturn's moon Titan. The internal structure of these bodies has been constrained from the measurements of the gravity coefficients from spacecraft radio tracking during close flybys (e.g. Galileo, Cassini) (*Anderson et al., 1996; Anderson et al., 2001; Iess et al., 2010, 2012*). The degree-2 gravity coefficients can then be used to infer the moment of inertia of the satellite, thus providing constraints on the density structure. The normalized moment of inertia about the spin axis, $C/(M_T R_T^2)$, is commonly calculated from the fluid Love number using the Radau-Darwin approximation. This formulation implies that the body's equipotential corresponds to the equilibrium shape due to fluid dynamic flattening. In this approximation, the degree-two gravity coefficients, $C_{20}$ and $C_{22}$, should be equal to -10/3 and are proportional to the fluid Love number.

In the case of Titan, the ratio between the observed $C_{20}^{obs}$ and $C_{22}^{obs}$ is close to -10/3 (*Iess et al. 2010*), suggesting that Titan is relatively close to hydrostatic equilibrium. We can therefore use the degree-two gravity coefficients to first order to estimate the fluid Love number and the corresponding moment of inertia factor. Depending on whether we use the $C_{20}$ or $C_{22}$ coefficients as a reference to determine the fluid Love number, the normalized moment of inertia varies between 0.335 and 0.342. For Ganymede and Callisto, the two coefficients have not been determined independently, and the -10/3 relationship has been prescribed when inverting the degree-two gravity coefficients. Using the degree-two coefficients, and applying the Radau-Darwin relationship, the moment of inertia factors of Ganymede and Callisto are estimated to be 0.31 and 0.355 (assuming hydrostatic equilibrium), respectively (*Anderson et al., 1996, 2001*). Titan's value is therefore intermediate between those of Ganymede and Callisto, suggesting a modest increase of density toward the center.

The high moment of inertia factor obtained by Galileo gravity measurements ($C/(MR^2) = 0.355$) (*Anderson et al., 2001*) suggests an incomplete ice-rock separation within Callisto. In contrast, Ganymede has a much smaller moment of inertia ($C/(MR^2) = 0.31$) (*Anderson et al., 2001*) and shows signs of past endogenic activity (*Pappalardo et al., 2004*). Hence, Ganymede is potentially fully differentiated with an icy upper mantle, a rocky lower mantle and a metallic core, which is the origin of a relatively intense intrinsic magnetic field (*Kivelson et al., 1998*). With similar size and mass, Titan may be an intermediate case between Callisto and Ganymede. Based on the estimation of its moment of inertia factor, Titan's interior is more differentiated than Callisto's but probably much less than Ganymede's interior. Like Callisto, Titan might still possess a layer of ice–rock mixture between a rocky core and an outer ice-rich mantle, unless the rocky core is composed mostly of hydrated minerals (*Sohl et al., 2010; Castillo-Rogez and Lunine, 2010*). A dense atmosphere is currently present at its surface but no intrinsic magnetic field has yet been detected on Titan. However the presence of a metallic core is plausible within Titan if enough iron was delivered during its accretion (*Grasset et al., 2000*). The fact that the interior of Callisto and possibly Titan may still contain a layer of ice–rock mixture suggests that the satellites have avoided significant melting during accretion and subsequent evolution.

We now discuss the apparently partially-differentiated state of Callisto and Titan, as suggested by their elevated moment of inertia factors (e.g. *Anderson et al., 2001; Iess et al., 2010; Gao and Stevenson, 2013*). In the absence of lithospheric strength and shell thickness variations, these satellites would adopt a perfectly hydrostatic shape, corresponding to a degree-two gravity field. On Titan, the existence of a non-negligible degree-three in the gravity field as well as significant topography suggest that non-hydrostatic effects may significantly affect the estimation of the moment of inertia factor (*Iess et al., 2010; Gao and Stevenson, 2013; Baland et al., 2014*) and that the MoI factor may be significantly smaller than the value estimated from the Radau-Darwin approximation. On Callisto, similar non-hydrostatic contributions originating in the lithosphere, which could not be tested with the Galileo spacecraft data, may also affect the estimation of its moment of inertia factor (*McKinnon, 1997; Gao and Stevenson, 2013*). On these two moons, the hydrostatic dynamical flattening is relatively small as they orbit relatively far from their respective planets, and therefore the non-hydrostatic contributions need to be correctly estimated in order to accurately infer the moment of inertia factor and the density profiles of their interior. Future measurements by the JUICE mission (*Grasset et al., 2013*) will allow us to test how far Ganymede and Callisto are from hydrostatic equilibrium.

*4.2 Differentiation process in water-dominated interiors*

Several scenarios have been proposed to explain the wide range of extents of differentiation among the icy moons that resulted from radioactive and accretional heating. *O'Rourke and Stevenson (2014)* showed that although rock–ice separation may be delayed by double-diffusive convection in the ice–rock interior, ice melting due to progressive radiogenic heating and subsequent differentiation cannot be prevented. Previous studies showed that it was possible to avoid melting if the accumulation of accretion energy was inefficient, i.e. if the energy was radiated away at a rate comparable to the accretion rate (e.g., *Schubert et al., 1981; Squyres et al., 1988; Kossacki and Leliwa-Kopystynski, 1993; Coradini et al., 1995; Grasset and Sotin, 1996*). Based on these models, the accretion timescales should be longer than 1 Myr to avoid significant melting and hence differentiation of Callisto, while an accretion timescale as short as $10^3$-$10^4$ yr may be possible for Ganymede. However, these timescales are dependent on the way heat deposition and cooling are treated. *Barr and Canup (2010)* proposed that the Ganymede–Callisto dichotomy can be explained through differences in the energy received during the Late Heavy Bombardment. Impacts would have been sufficiently energetic on Ganymede to lead to a complete ice-rock separation, but not on Callisto. More recently, *Monteux et al. (2014)* have developed a 3D numerical model that characterizes the thermal evolution of a satellite growing by multiple impacts, simulating the satellite growth and thermal evolution for a body radius ranging from 100 to 2000 km. Their results indicate that a satellite exceeding 2000 km in radius may accrete without experiencing significant melting only in case (i) its accretion history is dominated by small impactors (radius less than a few kilometers) and (ii) the conversion of impact energy into heat is unrealistically inefficient. In this context, they underlined that global melting for large bodies like Titan or Callisto is difficult to avoid.

Once the melting of the icy phase is initiated, ice-rock separation can proceed. This separation will lead to the accumulation of dense blocks of rock overlying a undifferentiated core consisting of a mixture of rock and ice (e.g., *Friedson and Stevenson, 1983, Kirk and Stevenson 1987*). The dense layer of accumulated rock is gravitationally unstable, which leads to large-scale ice-rock separation. *Friedson and Stevenson (1983)* suggested that an increase in the silicate volume fraction from 44% for Callisto to 55% for Ganymede could have induced an increase in the internal viscosity of between 1 and 2 orders of magnitude. This increase in the internal viscosity of Ganymede might have led to a failure of its solid-state convection self-regulation. This failure would then have limited heat loss, favored partial melting of the ice and enhanced a catastrophic runaway differentiation (*Kirk and Stevenson, 1987*). Similarly, *Nagel et al. (2004)* suggested that if Callisto accreted from a mixture of rock and ice and the average size of the rocks was of the order of meters to tens of meters, then the viscosity would range between $10^{13}$ and $10^{17}$ Pa s. For such viscosities, Callisto would have experienced a gradual, but incomplete unmixing of ice and rocks. This mantle/core separation potentially involves a large amount of heating due to the viscous dissipation of the potential energy (*Flasar and Birch, 1973*). Contrary to terrestrial planets (*Golabek et al., 2009*; *Ricard et al., 2009*), this runaway separation has not yet been modeled.

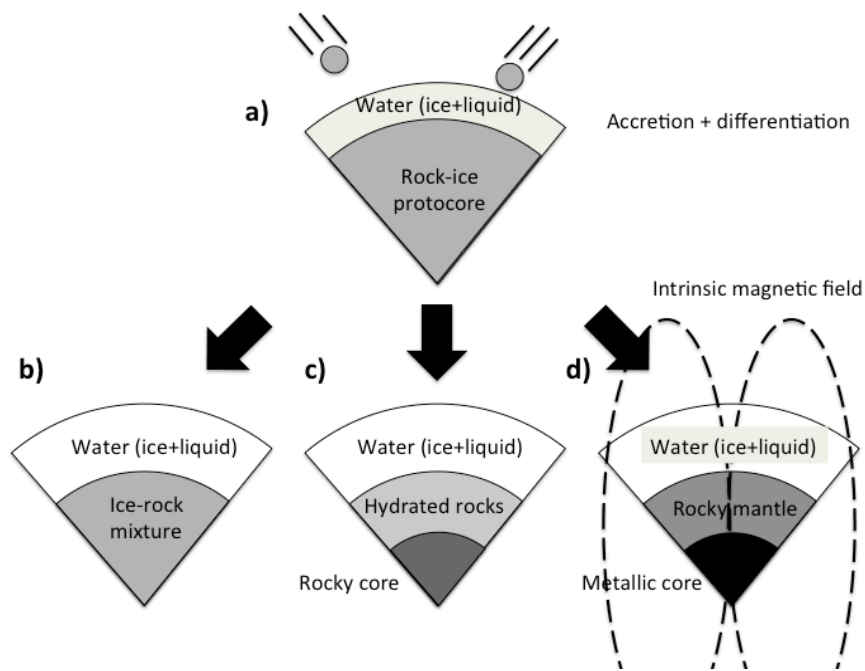

*Fig. 13: Schematic representation of the structure of an icy moon (a) just after accretion and after differentiation leading to (b) a Callisto-like structure (c) a Titan-like structure or (d) a Ganymede-like structure.*

*4.3 Metal-rock separation*

If a sufficient amount of native iron is present within the icy moon, an iron core may eventually form during or after the ice-rock separation. In this case, the deep interior would be comprised of a probably liquid iron core, and a convecting silicate mantle. Gravity measurements performed by the Radio

Science System onboard the Cassini spacecraft suggest that the deep interior of Titan is not fully differentiated (*Sotin et al., 2009a; Iess et al., 2010*). However, the presence of an iron core cannot be definitely ruled out, and additional measurements are required to determine whether Titan possesses an iron core like Ganymede. On Ganymede, a second differentiation event likely occurred during or after the ice-rock separation since enough heat had been stored in the metal-silicate mixture leading to the formation of an iron core. The major observable feature of this metal-silicate separation is the presence of an intrinsic magnetic field generated in its deep interior (*Kivelson et al., 1996*). While the ice-rock separation has been extensively studied, the modalities of metal-rock separation on icy moons and, as a consequence, the modalities of the generation of intrinsic magnetic fields remain poorly constrained. Another major difficulty is to constrain the composition and size of the metallic core. Depending on whether the core is pure iron or a sulfur-iron alloy, it could represent 1 to 30 % of the total mass of the icy moon (*Anderson et al., 1996*).

Although the interior of Ganymede is currently fully differentiated, the silicate rock and the iron likely coexisted and reacted with $H_2O$ (as ice or water) before metal/silicate separation occurred. In this context, it is difficult to envision the formation of a pure iron core because the oxidation of the iron would be unavoidable. This led *Crary and Bagenal (1998)* to suggest that $Fe_3O_4$-magnetite could maintain a strong remanent magnetic field within the rocky material of Ganymede. *Scott et al. (2002)* have shown that the presence of water could have enhanced the formation of low-density hydrated silicates and the alloying of iron with sulfur would have resulted in FeS-dominated cores. They performed high-pressure experiments that combined chondritic chemistry with water at a pressure of 1.5 GPa, temperatures between 300 °C and 800 °C and a range of oxygen fugacities. Their results indicate that sulfur may be a significant constituent of Ganymede's core but they did not observe the presence of either iron oxide species or metallic iron. Hence they concluded that magnetite is not a stable phase in the chondrite-water system in contrast with the suggestion of *Crary and Bagenal (1998)*.

Hence, Ganymede's core probably contains not just pure iron but also lighter elements. As mentioned above a candidate for the dominant core-forming assemblage is an alloy of iron (Fe) and sulfur (S) (*Hauck et al., 2006*). Indeed sulfur is an element that was abundant in the proto-solar system (*Hillgren et al., 2000*). Within Ganymede, the core sulfur concentrations could range between 12 and 20 wt %. Under the pressure conditions of Ganymede's core (≈6–10 GPa) a mixture of iron and sulfur forms a binary, eutectic system. A major consequence of adding sulfur to liquid iron is a depression of the liquidus (e.g., *Usselman, 1975; Fei et al., 1997*). Upon freezing, nearly pure iron (Fe) (*Li et al., 2001; Kamada et al., 2010*) crystallizes on the Fe-rich side of the eutectic and iron sulfide (FeS) on the Fe-poor side. Various experiments have revealed that the eutectic composition of Fe-FeS becomes increasingly Fe-rich with increasing pressure (e.g., *Chudinovskikh and Boehler, 2007; Fei et al., 1997, 2000; Morard et al., 2008*). *Buono and Walker (2011)* have suggested low and even negative gradients for the melting temperatures of Fe-rich Fe-FeS alloys in the pressure range of Ganymede's core. If the melting temperature has a lower gradient than the core temperature, iron will start to crystallize at the core-mantle boundary (CMB) and a growing snow zone will form during further cooling (*McKinnon, 1996; Hauck et al., 2006*). However, the unknown oxidation state of the interior during differentiation limits our ability to constrain whether the composition of a sulfur-bearing core is on the Fe- or FeS-rich side of the eutectic composition (e.g., *Scott et al., 2002; Breuer et al., 2015*). Another important element could be hydrogen. The interaction of native iron with hydrated material could have led to the addition of hydrogen to the iron core resulting in the presence of iron hydride (*Ohtani et al., 2005, Shibazaki et al., 2009*).

*4.4 Evidence for active and past aqueous processes*

A range of geophysical evidence indicates that several moons of Jupiter (Europa, Ganymede, Callisto) and Saturn (Titan, Enceladus, Mimas) harbor extensive liquid water reservoirs beneath their cold icy surfaces (*Khurana et al., 1998; Kivelson et al., 2000; Iess et al. 2014; Tajeddine et al., 2014; Thomas et al., 2016*). Detection of ammoniated phyllosilicates on Ceres (*De Sanctis et al., 2015*) as well as the young age of some extensional features observed on Pluto (*Hammond et al. 2016; Moore et al. 2016*) suggest the existence of a subsurface ammonia-rich ocean, at least in the past, inside these two dwarf planets. Based on evolution models, many other icy bodies are also predicted to harbor internal oceans underneath their cold icy shells (*Hussmann et al. 2006*). These massive oceans constitute the greatest volumes of potentially habitable space in the Solar System. If chemical interactions with the rocky core and the irradiated surface provide nutriments and energy sources in sufficient amounts, the development of primitive life may be envisaged in these water-rich worlds. Several observational clues suggest that complex interactions between the subsurface oceans and their warm rocky interiors are presently occurring.

The most convincing evidence for active aqueous processes is provided by the surprising observations of eruptions of water vapor and salted icy grains at Enceladus' active south pole, indicating connections with a subsurface ocean (*Postberg et al., 2009, 2011*). The discovery of silicon-rich nano-particles further indicates that hydrothermal interactions are currently occurring in porous warm (100-200 °C) rock and that hydrothermal products are quickly transported to the plume source (*Hsu et al., 2015; Sekine et al., 2015*). Moreover, analysis of gravity, shape and libration data shows that the ice shell thickness at the south pole is thinner than 5 km, with a globally average value of not more than

21 km, requiring about 25-30 GW of internal power for such a configuration to be thermally stable (*Cadek et al., 2016*). This provides additional constraints in favor of a warm rocky core, powered by strong tidal dissipation. However, it remains unclear how Enceladus reached such a hyperactive state.

On Europa, although there is no direct evidence yet, the occurrence of hydrothermal processes at the seafloor, at present or in recent geological history (< 100 Ma), is also likely. The observation of non-ice colored and hydrated materials in many disrupted areas on the surface of Europa points to exposure of oceanic species at the surface and to the possible existence of suboceanic hydrothermal systems (*Zolotov and Kargel, 2009*). The likelihood of presently active hydrothermal processes on Europa mostly depends on how much heat can be generated in the interior at present. Similarly to Enceladus and Io, tidal friction may contribute to Europa's internal budget, but in absence of observational constraints, it is still unclear if tidal friction in the rocky mantle is large enough at present to maintain active hydrothermal processes (*Sotin et al., 2009b*).

Despite the separation of the subsurface ocean from the rocky core by a thick high pressure ice mantle, water-rock interactions at the base of the thick ice mantle and the transport of brine through the ice mantle are still possible on the largest moons like Saturn's moon Titan and Jupiter's moon Ganymede (*Vance and Brown, 2013; Choblet et al., 2017*). It is very likely that water-rock interactions occurred on these large moons in the past, especially during the differentiation stage when a large volume of water percolated throughout rock-dominated layers toward the surface (*Tobie et al., 2013*) (Fig. 13). The detection of $^{40}$Ar, the decay product of $^{40}$K initially contained in silicate minerals, in Titan's atmosphere (*Niemann et al., 2010*) as well as the likely presence of salts in the internal ocean of Titan (*Béghin et al., 2012; Baland et al., 2014; Mitri et al., 2014*), Ganymede (*Saur et al., 2015*) and Callisto (*Khurana et al., 1998*) indicate that water-rock interaction occurred during the evolution of these large icy moons. However, it is still unclear for how long these water-rock interactions lasted and what effects they had on the thermo-chemical evolution of their interiors.

*4.5 Role of volatile compounds in the aqueous alteration within icy bodies.*

During the early evolution of Pluto, impact heating, radiogenic heat sources (short- and long-lived) and tidal heating due to the interaction with Charon have likely been significant to induce a complete differentiation of Pluto's interior (*Schubert et al., 2010, Sekine et al. 2017*). At present-day, because of the rapid decay of short-lived radioisotopes and since the Pluto/Charon system has reached the ''dual synchronous'' state in which Charon's orbital period, spin period, and Pluto's rotation period are equal, these two major heat sources are probably too weak to maintain a current water ocean within Pluto's interior (*Robuchon and Nimmo, 2011; Barr and Collins, 2015*). The surprising discovery of two quasi-circular mounds on Pluto by New Horizons, suspected to be of cryovolcanic origin (*Moore et al., 2016*), suggests that the interior of the dwarf planet is more active than initially anticipated and that water-volatile-rock interactions may have operated during a long period of time in its interior. The presence of significant amounts of anti-freeze compounds, such as ammonia and methanol, together with gas compounds in the form of clathrate hydrates may explain the occurrence of such activity on Pluto. This hypothesis is consistent with the detection of ammonia-rich and methane-rich areas on Charon (*Grundy et al., 2016*). Although ammonia has not been identified on Pluto's surface, that is dominated by $N_2$ ices (*Grundy et al. 2016*), its presence on Charon indicates that it may have played a key role in the chemical evolution of the Pluto-Charon system, as well as in those of other KBOs (*Hussmann et al., 2006*).

For decades, both ammonia and methanol, have been suspected to strongly affect the melting and crystallization process in icy bodies (e.g. *Lewis, 1971; Kargel, 1992*). The role of ammonia is expected to be particularly strong during the early stage in the evolution of icy bodies, when the first aqueous melts are generated during the post-accretional differentiation process and at the end of their evolution when the internal ocean becomes strongly concentrated in solutes by progressive crystallization of ice (*Hussmann et al., 2006; Malamud and Prialnik, 2015)*. Even in the absence of significant heat sources, the presence of ammonia, as well as methanol (both are present in comparable amount in comets (*Cochran et al., 2015*)) can sustain active aqueous processes for long periods of time. The recent detection of widespread ammoniated phyllosilicates across Ceres' surface (*De Sanctis et al., 2015*) may provide evidence for alteration of chondritic materials by N-bearing fluids. However, in absence of reliable experimental data on aqueous alteration by concentrated N-bearing fluids, it is difficult to investigate the conditions under which such ammoniated phyllosilicates may have formed.

**Conclusions and Outlook**

Water content and the internal evolution of terrestrial planets and icy bodies are closely linked. Indeed, the internal evolution and especially the formation time of the planetesimals and therefore the extent of radiogenic heating governs the amount of hydrous silicates and leftover rock-ice mixture available in the late stages of their evolution. In turn, water content may affect the early internal evolution of the planetesimals and metal-silicate separation by favoring the sinking of metal diapirs over percolation processes. On the other hand, the absence of water could change the descent mode on larger terrestrial bodies from sinking diapirs to diking, thus affecting the heat partitioning between iron and silicates and modifying the thermal state of the accreting planet. Moreover, water content can contribute to an increase of $f_{O2}$ and thus affect the concentration of siderophile elements within the silicated reservoirs of the Solar System objects. Finally, the water content strongly influences the differentiation rate among the icy moons, controls their internal evolution and governs the alteration processes occurring in their deep interiors.

Partitioning of water between the mantle and the surface/atmosphere is also a key factor in regulating both planetary climate and internal dynamics. This partitioning is controlled by release of water by erupting melts, by hydrothermal alterations, mainly serpentinization, trapping some water in the form of hydrated minerals in the crust, and by recycling part of it deep into the mantle where it can be dissolved in nominally-anhydrous minerals. On Earth, where the water cycle is driven by plate tectonics, exactly how water partitions between surface and interior remains uncertain, but it is commonly thought to be of the order of 1. Parameterized models have recently been proposed to address the question of water cycle on other terrestrial planets, with different sizes and water contents (Crowley et al., 2011; Cowan and Abbot 2014; Schaeffer and Sasselov, 2015). These parameterized approaches enable some important characteristics of the water cycle to be understood, especially how surface gravity and the water volume at the surface influence the interior/surface exchange. However, they cannot describe the complexity of water transport in the mantle and its feedback on internal melting and crustal recycling. Describing correctly the coupled evolution of surface and internal reservoirs of water requires the use of thermo-chemical convection codes in 3D or at least 2D, including exchange of water between the crust and mantle, the convective transport in the mantle, its retroaction on viscosity, melting and associated outgassing.

Finally, with the discoveries of thousands of exoplanets, we are now entering a new era in comparative planetology. Even if the dataset is limited to the mass, radius and orbital characteristics of the planetary systems, it already offers a unique opportunity to test the models of planet formation and evolution. Model mass-radius relationships indicate a great diversity of interior compositions and atmospheric extents for the Super-Earth/Mini-Neptune-planet class (e.g. Howard 2012), suggesting a wide range of volatile contents and compositions. Hence, planetary worlds are probably much more diverse than originally thought, with a wide range of water and other volatile contents.

**Acknowledgements**. We thank L. Elkins-Tanton, A. Morbidelli and two anonymous reviewers for detailed and thoughtful comments that helped to improve the manuscript. D.C.R. was supported by the European Research Council Advanced Grant "ACCRETE" (contract number 290568) and additional support was provided by the German Science Foundation (DFG) Priority Programme SPP1833 "Building a Habitable Earth" (Ru 1323/10-1). J.M. was funded by the Auvergne Fellowship program. This is Laboratory of Excellence ClerVolc contribution no. XX. G.J.G. thanks Taras Gerya for providing the code I2MART. G.T acknowledges support from the project ANR OASIS.